\documentclass[twocolumn,showpacs,preprintnumbers,amsmath,amssymb]{revtex4}
\usepackage{graphicx}
\usepackage{dcolumn}
\newcommand{\beq}{\begin{equation}}
\newcommand{\eeq}{\end{equation}}

\begin{document}


\title{Nonlocal phenomenology for anisotropic MHD turbulence}


\author{A. Alexakis}
\affiliation{Laboratoire Cassiop\'ee, Observatoire de la C\^ote d'Azur, BP 4229, Nice Cedex 04, France}


\begin{abstract}
A non-local cascade model for anisotropic MHD turbulence in the presence of a
guiding magnetic field is proposed. The model takes into
account that (a) energy cascades in an anisotropic manner
and as a result a different estimate for the cascade rate
in the direction parallel and perpendicular to the guiding field is made.
(b) the interactions that result in the cascade are between different
scales. Eddies with wave numbers $k_\|$ and $k_\perp$ interact with eddies with
wave numbers $q_\|,q_\perp$ such that a resonance condition between the wave numbers
$q_\|,q_\perp$ and $k_\|,k_\perp$ holds.
As a consequence energy from the eddy  with wave numbers $k_\|$
and $k_\perp$ cascades due to interactions with eddies
located in the resonant manifold whose wavenumbers are determined by:
$q_\|\simeq \epsilon^{{1}/{3}}k_\perp^{2/3}/B$, $q_\perp=k_\perp$ and
energy will cascade along the lines
$k_\|\sim C+k_\perp^{2/3} \epsilon^{1/3}/B_0$.
For a uniform energy injection rate in the parallel direction the resulting energy spectrum is
$E(k_\|,k_\perp)\simeq \epsilon^{2/3}k_\|^{-1}k_\perp^{-5/3}$. For a
general forcing however the model suggests a non-universal behavior.
The connections with previous models, numerical simulations and weak turbulence theory are discussed.
\end{abstract}




\maketitle  

\section{Introduction}

Magnetic fields are met very often in astrophysics; interstellar
medium, accretion discs, the interior of stars and planets. In most
of these cases  the magnetic fields are strong enough to play a
dynamical role in the evolution of the involved astronomical objects
\citep{Zeldovich}. Furthermore, the fluid and magnetic Reynolds
numbers are large enough so that a large number of scales are exited
and coupled together, making it very difficult to calculate the
evolution of these systems even with the power of present day
computers. As a result a turbulence theory that models the behavior
of the small unresolved scales is in need. The simplest set of equations 
that describes the evolution of the flow and the magnetic field when the
two are coupled together are the magneto-hydro-dynamic (MHD)
equations that in the Els\"asser formulation are written as
\begin{equation}
\label{MHD}
\partial_t {\bf z}^\pm = \pm {\bf B} \cdot \nabla {\bf z}^\pm - {\bf z}^\mp  \cdot \nabla {\bf z}^\pm -\nabla P + \nu \nabla^2 {\bf z}^\pm
\end{equation}
where ${\bf z}^\pm={\bf u\pm b}$ with ${\bf u}$ the velocity,
${\bf b}$ the magnetic field, $\nu$  the molecular viscosity
assumed here equal to the magnetic diffusivity $\eta=\nu$, $B$ is a
uniform magnetic field, and incompressibility $\nabla \cdot z^\pm=0$
has been assumed.

For zero viscosity the above equations conserve three quadratic
invariants the magnetic Helicity (that we are not going to be
concerned with in the present work) and the two energies
$E^{\pm}=\int (z^\pm)^2 dx^3$. The question then arises in the limit
of infinite Reynolds number is there a physical process under which
the two energies cascade to sufficiently small scales so that
they can be dissipated?


For hydrodynamic turbulence a description of such a process
exists and was given by
\citet{Kolmogorov1941} (K41). In his phenomenological description the energy
$z_l^2$ at a scale $l$ interacts with similar size eddies and
cascades in a timescale $l/z_l$. As a result in a
statistically steady state the energy cascades in a scale
independent way at a rate $\epsilon \simeq z_l^3/l$ that leads to
the prediction $z_l\sim l^{1/3}$ or in terms of the 1-D energy
spectrum $E(k)\sim k^{-5/3}$. Since the phenomenological
description of the energy cascade in hydrodynamic turbulence
there have been attempts to derive similar results for
MHD flows. However, non-trivial difficulties arise when a mean
magnetic field is present.
First the MHD equations are no longer
isotropic resulting in an anisotropic energy flux and energy spectrum.
Simple dimensional arguments can not be used
to estimate the degree of anisotropy that is a dimensionless quantity.
Second the MHD equations are
no longer scale invariant, as a result simple power law behavior of
the energy spectrum is expected only in the small or large B limit
that scale similarity is recovered.
Finally, it is not clear that
interactions of similar size eddies (local interactions)
dominate the cascade. Different size eddies could
play an important in cascading the energy.

The first model for MHD turbulence was proposed by \citet{Iroshnikov1963} and by
\citet{Kraichnan1965} (IK). The IK-model assumes isotropy and that the
time scale of the interactions of two wave packets of size $l$ is
given by the Alfven-time scale $\tau_{_A}\sim l/B$. The energy
cascade due to a single collision is given by $\Delta z^2 \sim
(z^3/l) \tau_{_A}$. The number of random collisions that would be
required then to cascade the energy is going to be $N\sim (z^2 /\Delta
z^2)^2$. As a result the energy will cascade in a rate $\epsilon
\sim E/N \tau_{_A} \sim z^4/(Bl)$ and therefore $z_l \sim (\epsilon
B)^{1/4}l^{1/4}$. The resulting 1-D energy spectrum is then given
by $E(k)\sim (\epsilon B)^{1/2} k^{-3/2}$.
The assumption of
isotropy however has been criticized in the literature and
anisotropic models have been proposed for the energy spectrum.
\citet{Goldreich1995} (GS) proposed that in strong turbulence the
cascade happens for eddies such that the Alfven time scale
$\tau_{_A}\sim Bk_\|$ is of the same order with the non-linear time
scale $\tau_{_{NL}}\sim z k_\perp$ (so called critical balance
relation), where $k_\|$ and $k_\perp$ are the parallel and
perpendicular to the mean magnetic field wavenumbers respectively.
Repeating the Kolmogorov arguments then one ends up with the
energy spectrum
$E(k_\|,k_\perp)\sim k_\perp^{-5/3}$ with
the parallel and perpendicular wave numbers following the relation
$k_\|\sim k_\perp^{2/3}$.
A generalization of this result was proposed by \citet{Galtier2005} where
the ratio of the two time scales $\tau_{_A}/\tau_{_{NL}}$ was kept
fixed but not necessarily of order one, in an attempt to model MHD
turbulence both in the weak and the strong limit.
\citet{Bhatta2001} (BN) repeated the IK-model arguments replacing
the nonlinear time scale by $\tau_{_{NL}}\sim l_\perp/z_l$ and the
Alfven time scale by $\tau_{_{A}}\sim l_\|/B$.
Further assuming that the
cascade is happening only in the $k_\perp$ direction
obtained the energy spectrum  $E(k)\sim (\epsilon B)^{1/2} k_\perp^{-2}k_\|^{-1}$.
Finally \citet{Zhou2004} (ZMD) suggested using as
time scale the one given by the inverse average of the Alfven and
nonliner time scale
$\tau^{-1}=(\tau_{_A})^{-1}+\tau_{_{NL}}^{-1}$ to obtain a smooth
transition from the K41 to the IK and the anisotropic BN result
depending on the amplitude of the guiding field.

Although this large variety of models exists the agreement with
observations \citep{Goldstein1995} and with the results of numerical
simulations
\citep{Muller2005,Ng2003,Maron2001,Cho2000,Biskamp2000,Ng1996}
is only partially satisfactory and seems to be case dependent.
Furthermore, all these models assume locality of interactions (i.e.
only similar size eddies interact and only one
length scale is needed in each direction in the phenomenological description).
Locality of interactions however has been shown to be in question by
both theoretical arguments and analysis of data in numerical
simulations even in the isotropic case
\citep{Alexakis2005,Debliquy2005,Yousef2007}. Furthermore the
rigorous result for weak turbulence \citep{Galtier2000,Galtier2002}
has shown that only modes in the resonant manifold $k_\|=0$ are
responsible for the energy cascade. It seams reasonable therefore
that non-locality is an essential ingredient of MHD turbulence that
needs to be taken in to account in a model. 
In addition we expect that the energy will not cascade isotropically
so not only the amplitude of the energy cascade rate is of
importance but also the direction.
With these two points in mind (anisotropy and non-locality)
we try to construct a non-local
model, for the energy cascade.


To begin with the new model let us consider a MHD flow
in a statistically steady turbulent state forced at
large scales, in the presence of mean magnetic field $B$.
We will denote the
two 2-D energy spectra as $E^+(k_\perp, k_\|)$ and $E^-(k_\perp,
k_\|)$ where the total energy is given by $E_{_T}^\pm \equiv \int
E^\pm dk_\perp dk_\|$.
%
%
For simplicity assume $E^-(k_\perp, k_\|)\sim E^+(k_\perp, k_\|)$ (i.e.
negligible cross helicity) and drop the $\pm$ indexes leaving the case
$E^-\ll E^+$ to be investigated in the future. To shorten the
notation we will write $E_k=E(k_\perp, k_\|)$.
The index $k$ denotes that $E_k$ depends on the
wavenumbers $k_\|$ and $k_\perp$.

\begin{figure}
\includegraphics[width=8cm]{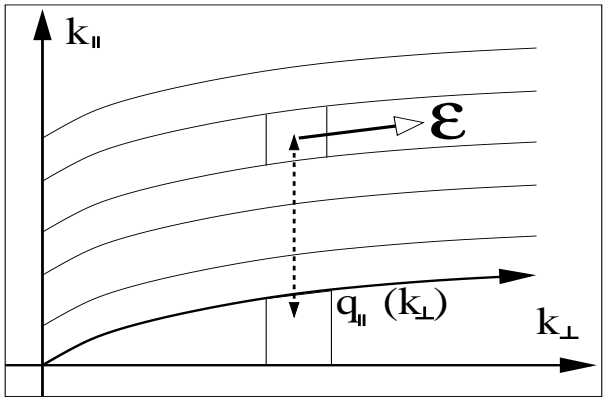}
\caption{An illustration of the non local model. A eddy at $(k_\|,k_\perp)$ interacts (non-locally)
with an eddy at the resonant manifold $(q_\|(k_\perp),k_\perp)$ cascading the energy in the direction of the arrow.
\label{fig1}}
\end{figure}

Let us now consider two eddies of different scales $z^+_k$ and $z^-_q$ interacting.
Let us assume that the $z^+_k$ eddy has  wave numbers $\sim k_\perp,k_\|$ and
the  $z^-_q$ eddy has wave numbers $\sim q_\perp,q_\|$.
Here we will focus on the cascade of the energy of the $z^+_k$ eddy,
the cascade of the $z^-_q$ can be obtained by changing the indexes $k,q$ and $\pm$.
From the form of the non-linear term we expect by  dimensional analysis
that the rate of energy cascade of the $z^+_k$ eddy will be
\begin{eqnarray}
\mathcal{E}(k)  & \sim (z_k^+)^2 (z_q^-) q \qquad \qquad \qquad \nonumber \\
                & \sim [k_\perp k_\| E_k ]
                     [ q_\perp q_\| E_q]^{1/2} q
\end{eqnarray}
and similar for the energy cascade rate of the $z^-_q$ eddy.
Note that in such an interaction
the energy will not cascade isotropically but it will depend on the value of ${\bf q}$.
In a interaction of the two eddies the energy of the $z^+_k$ eddy will move from the wavenumber 
${\bf k}$ to the wave number ${\bf k+q}$.
If $q_\|\ll q_\perp$ then most of the cascade will be in the $q_\perp$ direction.
As a result we need to define separately the rate energy cascades
to larger $k_\perp$: $\mathcal{E}_\perp (k)$  and the rate energy cascades to larger
          $k_\|$:    $\mathcal{E}_\|    (k)$ as:
\beq
\label{Eperp}
\mathcal{E}_\perp \sim  [ k_\perp k_\| E_k ] [ q_\perp q_\| E_q ]^{1/2} q_\perp
\eeq
\beq
\label{Epara}
\mathcal{E}_\|    \sim  [ k_\perp k_\| E_k ] [ q_\perp q_\| E_q ]^{1/2}  q_\|.
\eeq
Note that in writing the equations above we have not taken in to account possible
scale dependent correlations between the two fields that could reduce the energy
cascade. Such an effect has been taken into account by \citet{Boldyrev2005} based
on the GS model and could be incorporated in the present model.
However we will not make such an attempt here since we want to present the model
in its simplest form.
Equations \ref{Eperp} and \ref{Epara} express the rate energy cascades in the absence
of a mean magnetic field and are valid only when $|q|<|k|$ because small eddies
although they have a stronger shear rate $z_qq$ decorrelate making them less effective in cascading the energy.
However in the presence of guiding field not all wave numbers $q$
are as effective in cascading the energy $E_k$.
Because the two eddies $z^+_k$,$z^-_q$  travel in opposite directions the time they will interact
will be the  Alfven time $\tau_{_A} \sim [q_\| B_0]^{-1}$ and the time needed to cascade the
energy will be $\tau_{_{NL}} \sim (z^-|q|)^{-1} \sim [ |q| \sqrt{q_\perp q_\| E_q}]^{-1}$.
Therefore, from all the available wavenumbers only
the wave numbers with $\tau_{_A} \gtrsim \tau_{_{NL}}$ will be effective
in cascading the energy.
This restriction leads to:
\beq
\label{qpara1}
q_\| B \lesssim   q_\perp \sqrt{q_\perp q_\| E_q}
\eeq
where we used the approximation $|q| \simeq q_\perp$ as a first order approximation
of $|q|$ for large $B$.
This relation looks very similar to the critical balance relation of
the GS model. However in this case the relation \ref{qpara1} gives
the wave numbers that the eddy $z^+_k$ will interact with and does not
restrict the location of $z^+_k$ in spectral space.
We are going to refer to the set of wavenumbers that satisfy the relation above as
the resonant manifold and
we are going to use this  relation as an equality because from the allowed wavenumbers $q_\|$
the ones closer to $k_\|$ will be more effective in cascading the energy.
Here as in the equations \ref{Eperp} and \ref{Epara}, we assume that $q_\|<k_\|$.
Finally since the mean magnetic field does not directly effect the $\perp$ direction we assume that
similar size $k_\perp$ and $q_\perp$ are the most effective to cascade the energy
$q_\perp \sim k_\perp$ ({i.e. locality in the $k_\perp$ direction}).
So equation \ref{qpara1} is written as
\beq
\label{qpara}
q_\|\simeq \frac{ k_\perp^3 E_q}{B_0^2}
\eeq
We are now ready to impose the constant energy flux condition that would lead to a stationery
spectrum. Because the cascade is anisotropic, constant energy flux now reads:
\beq
\label{flux}
\partial_{k_\perp} \mathcal{E}_\perp + \partial_{k_\|} \mathcal{E}_\| =0
\eeq
The equations  \ref{Eperp},\ref{Epara},\ref{qpara},\ref{flux} form the basic equations of our model.
It is worth noting, that in this model the cascade of energy decreases by the introduction of the guiding field
not because the individual interactions weakened but because the number of modes that are able to cascade the energy
decreases due to the resonance condition \ref{qpara}.
A sketch of the mechanisms involved in the model is shown in figure \ref{fig1}.


First let us concider the weak turbulence limit that is obtained
in the limit $B\to \infty$.
For large $B$ based on equation \ref{qpara} the resonant manifold becomes very thin $q_\|/k_\perp \ll 1$.
Furthermore because $\mathcal{E}_\|/\mathcal{E}_\perp \sim q_\|/k_\perp \ll 1$
we can neglect the cascade in the parallel direction.
If also $E_k$ is non-singular at $k_\|=0$ we have that $E_q\simeq E(k_\perp,0)$.
Substituting $q_\|$ from \ref{qpara} in \ref{Eperp} and imposing the constant flux condition \ref{flux}
we obtain:
\beq
\mathcal{E}_\perp \sim  k_\perp^4  k_\| E_k E(k_\perp, 0 )/B=\epsilon(k_\|).
\eeq
Note that the spectrum $E(k_\perp, k_\|)$ depends on the energy of the resonant
manifold $E(k_\perp, 0 )$ just like the weak turbulence result and
unlike what the BN local theory for weak turbulence predicts.
If the energy spectrum for $E(k_\perp, k_\|)$ and for the resonant manifold $E(k_\perp, 0)$
scale like $k_\perp^n$ and $k_\perp^m$ respectively then we end up with
the weak turbulence prediction \citep{Galtier2000}:
\beq
\label{WC}
m+n=-4
\eeq
Assuming that the two spectra are
 smooth around $k_\|=0$ (just like the weak turbulence theory
needs to assume) we obtain:
\beq
\label{Wspec}
E^\pm \sim k_\perp^{-2} \sqrt{B\epsilon(k_\|)/k_\|}.
\eeq

As we decrease the value of $B$ we need to take in to account that the energy
cascade in the $\|$ direction is non zero. In this case energy does not cascade
in the $\perp$ direction but cascades along the lines that
are tangent to the direction of $\mathcal{E}$  and
satisfy $dk_\|/dk_\perp=\mathcal{E}_\perp/\mathcal{E}_\|=q_\|/k_\perp$ or
\beq
\label{lines}
k_\|=\int_0^{k_\perp} \frac{q_\|(k_\perp')}{k_\perp'}dk_\perp'+C.
\eeq
Let $\lambda$ be the length along such a curve; then we can move
to a new coordinate system given by ($\lambda,C$).
In this new coordinate system the constant flux relation to first order in $q_\|$ reads
\beq
\frac{d|\mathcal{E}|}{d\lambda}=\frac{d}{d\lambda}\left(  [ k_\perp k_\| E_k ] [ q_\perp q_\| E_q ]^{1/2} |q| \right)=0
\eeq
or
\beq
\label{energyGS}
[ k_\perp k_\| E_k ] [ k_\perp q_\| E_q ]^{1/2} k_\perp = \epsilon(C)
\eeq
where
only terms up to order $q_\|$ are kept.
Letting $k_\|\to q_\|$  we obtain the equation for the resonant manifold.
\beq
[ k_\perp q_\| E_q ] [ k_\perp q_\| E_q ]^{1/2} k_\perp = \epsilon(0) \equiv \epsilon_0.
\eeq
Substituting $q_\|$ from \ref{qpara} and solving for $E_q$ we obtain
\beq
\label{EnR}
E_q =E(q_\|(k_\perp),k_\perp)= \epsilon_0^{1/3} B k_\perp^{-7/3}.
\eeq
The equations \ref{qpara} and \ref{lines} then gives us
\beq
q_\|= k_\perp^{2/3} \epsilon_0^{1/3}/B \,\, \mathrm{and}\,\, k_\|=\frac{3}{2}k_\perp^{2/3} \epsilon_0^{1/3}/B+C.
\eeq
Returning to the equation for the energy energy \ref{energyGS} we get
\beq
\label{spec1}
E_k =  \frac{ \epsilon(C) \epsilon_0^{-1/3}}{ k_\perp^{5/3} k_\|}
\eeq
where $C$ is given by equation \ref{lines} and
the predicted spectra \ref{spec1},\ref{spec2} are valid in the range $q_\|< k_\| <\infty$.
For smaller values of $k_\|$ the condition $k_\|< q_\|$ that we initially assumed is not satisfied.
The energy of the modes inside the resonant manifold is given by \ref{EnR} and no singularity
at $k_\|=0$ exists.

In the special case that $\epsilon(C)$ is a constant $\epsilon(C)=\epsilon_0$
that corresponds to a uniform injection rate per unit of wavenumber ($k_\|$)
at the large scales $k_\perp \to 0$
the spectrum reduces to
\beq
\label{spec2}
E_k =  \frac{ \epsilon_0^{2/3} }{ k_\perp^{-5/3} k_\|}
\eeq
but in general the spectrum will depend on the way energy is injected in the system.
The non-universality that the model suggests is due to the fact that we assumed
that the energy cascades in a deterministic way only along the lines in the ($k_\|,k_\perp$) plane
given by \ref{lines}. In reality energy will not cascade strictly along the lines \ref{lines}
but there is going to be some exchange of energy between lines that could bring the energy spectrum in the form
\ref{spec2}. However, if and how fast a universal spectrum can be obtained in MHD 
is not an easy question to answer.
This question is related to the return to isotropy of an
anisotropicaly forced flow in hydrodynamic turbulence,  
that is still an open question.
If indeed in MHD turbulence in the presence of a guiding field there is a universal
spectrum this is expected to happen at smaller scales than in hydrodynamic turbulence
because nonlinear interactions are weaker.

So far we concerned ourselves with only large values of $B$.
In principle we could extend our results to smaller values of $B$
with out making some of the approximations used to arrive at the results \ref{spec1},\ref{spec2}.
However, such procedure leads to more complex equations that prevent us from deriving
the energy spectrum in a compact form and we do not make such an attempt at present.

It is worth emphasizing the similarities the current model has with GS model.
Both models emphasize the role
of the manifold $k_\|\simeq k_\perp^{2/3}\epsilon_0^{1/3}/B$,
obtained by the resonance condition \ref{qpara} in out model
or the critical balance condition in the GS model.
However in this model
the cascade is not restricted in this manifold, but instead all modes in the ($k_\|,k_\perp$) plane cascade
due to non-local interactions with the modes in this manifold.

An other point we need to emphasize is that taking $B\to \infty$ does not reduce
the predicted spectra in \ref{spec1} to the weak turbulence limit \ref{Wspec}.
This is due to the two  different limiting procedures followed.
In the first case (eq.\ref{Wspec}) first the limit $B\to\infty$ was taken and then the 1-D flux was determined,
while in the second case \ref{spec1} we first obtained the 2-D energy flux and then the limit $B\to\infty$ was taken.
Note however that the condition \ref{WC} is satisfied in both cases and the spectrum
is also smooth at $k_\|=q_\|$ because although the resonant manifold scales
like $k_\perp^{-7/3}$ while the rest of the spectrum scales like $k_\perp^{-5/3}k_\|^{-1}$
the resonant manifold widens as $k_\perp$ increases.
It is possible as we discuss bellow that in different (numerical) setups
either of the two limiting procedures
can be valid and different spectra could be obtained.

In numerical simulations a finite discrete number of modes is kept.
Based on this model
the cascade rate is reduced in the presence of a mean magnetic field
not because the individual interactions themselves are weakened but because
the number of modes that that interact effectively is reduced.
If the modes in a numerical simulation with the smallest non-zero wavenumber $k_1=2\pi/L$
(where $L$ is the box height) is larger than the resonant manifold
( $k_1>q_\|$ ) then if $B$ is further increased
the scaling with $B$ of the energy dissipation rate
will be lost and the spectrum exponents could change,
since the number of modes in the
resonant manifold already have taken their minimum value (i.e. the number of
modes that have $k_\|=0$).
Furthermore a difference in the energy spectrum exponents can be expected
in numerical simulations if the modes inside the resonant manifold are not forced.
The sensitivity of the model to the way the system is forced could in part
explain the disagreement in the measured spectrum exponents.



The author is grateful for the support 
he received from
the Observatoire de la C\^ote d'Azur.




\begin{thebibliography}{}
\bibitem[Alexakis, Mininni \& Pouquet (2005)]{Alexakis2005}
         Alexakis A., Mininni P.D. \& Pouquet A.,
         2005, Phys. Rev. E 72, 046301


\bibitem[Boldyrev (2005)]{Boldyrev2005}
    Boldyrev, S.,
    2001, \apj, 626, L37-L40.


\bibitem[Bhattacharjee \& Ng (2001)]{Bhatta2001}
       Bhattacharjee, A., Ng, C. S.,
       2001, \apj, 548, 318-322

\bibitem[Biskamp \& M{\"u}ller (2000)]{Biskamp2000}
         Biskamp, D. \& M{\"u}ller, W.C.,
         2000 Phys. Plasmas, 7, 4889

\bibitem[Cho \& Vishniac (2000)]{Cho2000}
         Cho J. \& Vishniac E.T., 2000 \apj 539, 273

\bibitem[Debliquy, Verma \& Carati (2005)]{Debliquy2005}
         Debliquy O.,Verma M.K. \&  Carati D.,
         2005, Phys. Plasmas 12, 042309

\bibitem[Galtier et al.(2005)]{Galtier2005}
        Galtier, S., Pouquet, A., \& Mangeney A.,
        2005, Phys. Plasmas 12, 092310

\bibitem[Galtier et al.(2002)]{Galtier2002}
        Galtier, S.; Nazarenko, S. V.; Newell, A. C.; Pouquet, A.,
        2002, \apj, 564, L49-L52.

\bibitem[Galtier et al.(2000)]{Galtier2000}
        Galtier, S.; Nazarenko, S. V.; Newell, A. C.; Pouquet, A.
        2000, J. Plasma Phys., 63, 447.

\bibitem[Goldreich \& Sridhar (1995)]{Goldreich1995}
         Goldreich, P., Sridhar, S. 1995, \apj, 438, 763.

\bibitem[Goldstein, Roberts \& Matthaeus]{Goldstein1995}
         Goldstein, M.L., Roberts D.A., \& Matthaeus W.H., 1995,
         Annu. Rev. Astron. Astrophys. 33, 283.

\bibitem[Iroshnikov (1963)]{Iroshnikov1963}
         Iroshnikov, P. 1963, Soviet Astron., 7, 566.

\bibitem[Kolmogorov (1941)]{Kolmogorov1941}
         Kolmogorov, A.N. 1941,
         Dokl. Akad. Nauk SSSR 30,299

\bibitem[Kraichnan (1965)]{Kraichnan1965}
         Kraichnan, R. 1965, Phys. Fluids, 8, 1385

\bibitem[Maron \& Goldreich (2001)]{Maron2001}
         Maron J., Goldreich P., 2001 \apj 554, 1175

\bibitem[M{\"u}ller \& Grappin (2005)]{Muller2005}
         M{\"u}ller W.C., Grappin, R., 2005, Phys. Rev. Lett. 95, 114502


\bibitem[Ng \& Bhattacharjee (1996)]{Ng1996}
         Ng, C.S.,  Bhattacharjee, A. 1996, \apj, 465, 845.


\bibitem[Ng, Bhattacharjee \& Germaschewski (2003)]{Ng2003}
         Ng, C.S.,  Bhattacharjee, A., Germaschewski K., Galtier, S.  2003, Phys. Plasmas, 10, 1954


\bibitem[Yousef, Rincon \& Schekochihin (2007)]{Yousef2007}
         Yousef T.A., Rincon F., \& Schekochihin A.A., 2007, J. Fluid Mech. 575, 111.

\bibitem[Zeldovich, Ruzmaikin \& Sokoloff (1990)]{Zeldovich}
         Zeldovich Ya. B., Ruzmaikin A.A., Sokoloff D.D.
         ``Magnetic Fields in Astrophysics" 1990 Gordon \& Breach Science Pub.


\bibitem[Zhou, Matthaeus \& Dmitruk (2004)]{Zhou2004}
         Zhou Y., Matthaeus W.H., Dmitruk P. 2004
         Rev. Mod. Phys., 76, 1015-1035




\end{thebibliography}
\end{document}